\documentclass[prx, amsmath, amssymb, reprint]{revtex4-2}
\usepackage{comment}
\usepackage{amsmath}
\usepackage[nolist]{acronym}
\usepackage{graphicx}
\usepackage[caption=false]{subfig}
\usepackage{soul}
\usepackage{placeins}
\usepackage[colorlinks=true, linkcolor=black, citecolor=black, urlcolor=blue]{hyperref}

\begin{document}

\title{Impact of ion-electron collisions on nonlocal ion heat conduction, viscous stress, and diffusion}
\author{Nicholas Mitchell$^{1,2}$}
\author{David Chapman$^2$}
\author{Grigory Kagan$^1$}%
\affiliation{$^1$The Blackett Laboratory, Imperial College, London SW7 2AZ, UK
}%
\affiliation{$^2$First Light Fusion Ltd., Unit 9/10 Oxford Pioneer Park, Mead Road, Yarnton, Kidlington OX5 1QU, UK}

\date{\today}% It is always \today, today,

\begin{abstract}
     By applying a first-principles reduced kinetic method, this work demonstrates the impact of ion-electron collisions on ion transport for strongly inhomogeneous plasmas in the nonlocal regime, where collisionality is insufficient to enforce local thermal equilibrium due to sharp gradients. Ion heat conduction and viscous stress in both unmagnetized and magnetized plasmas are considered, as well as inter-species diffusion in multi-species plasmas. Most notably, even for equal ion and electron temperatures, ion-electron collisions are found to substantially modify the peak nonlocal heat flow, whereas nonlocal preheats are strongly suppressed since streaming suprathermal particles are further out in the tail of the cold ion distribution where ion-electron collisions become dominant.
\end{abstract}
%Analytic work by Ji and Held has shown that ion-electron collisions have a significant influence on the ion heat conductivity and viscosity in a simple plasma; however, these closures are limited to the highly collisional regime where transport is local, and their extension to arbitrary collisionality is limited to parallel transport in a near-homogeneous plasma.

%Heat transport, viscosity, and diffusion are key mechanisms that dictate the behavior of plasmas, with wide significance to both astrophysics and confined fusion. In the highly collisional limit where transport is local, analytic work by Ji and Held \cite{ji_closure_2013,ji_ion_2015} has shown that ion-electron collisions have a significant influence on the ion heat conductivity and viscosity in a simple plasma. This work demonstrates that ion-electron collisions also impact ion heat conduction, viscous stress, and diffusion in the nonlocal regime, where gradients are sufficiently sharp such that collisionality is insufficient to enforce locality of transport. Even for equal ion and electron temperatures, ion-electron collisions notably modify the peak nonlocal heat flow similarly to the local limit, whereas nonlocal preheats are strongly suppressed since streaming suprathermal particles are further out in the tail of the cold ion distribution where ion-electron collisions become dominant. Ion-electron collisions also notably influence the nonlocal ion viscous stress and multi-species transport fluxes.

\maketitle

\begin{acronym}
\acro{VFP}{Vlasov-Fokker-Planck}
\acro{MFP}{mean free path}
\acro{BGK}{Bhatnagar–Gross–Krook}
\acro{DF}{distribution function}
\acro{DDF}{deviation of distribution function from Maxwellian}
\acro{CoM}{centre-of-momentum}
\acro{RKM}{reduced kinetic method}
\acro{LTE}{local thermal equilibrium}
\acro{ICF}{inertial confinement fusion}
\acro{MD}{molecular dynamics}
\end{acronym}

%\begin{enumerate}
%    \item Heat transport important for plasmas
%    \item Usually electron conduction, but ion conduction can be comparable or dominant due to magnetic fields
%    \item Nonlocal effects usually viewed for electrons, but can be treated for ions and have been in previous work
%    \item Ion viscosity well understood to be significant, and nonlocal treatment given in previous work
%    \item Ion-electron collisions shown to significantly impact local ion transport by Ji and Held, so should impact nonlocal ion transport
%    \item This work uses RKM to investigate these effects
%\end{enumerate}

\section{Introduction}

Collisional transport of heat, momentum, and impurities is widely appreciated to be of fundamental significance to understanding plasma dynamics and is of great practical importance to fusion for energy production \cite{icf_kinetics_review,rosenberg_assessment_2015,osti_22489850}. Braginskii's 1965 monograph laid the foundation for fluid and transport theory for a simple magnetized plasma \cite{Braginskii1965ReviewsOP}, providing local analytic closures for collisional transport quantities including the ion heat flux and viscous stress tensor. In deriving the ion transport coefficients appearing in these local closures, Braginskii neglected ion-electron collisions since electrons are much lighter than ions. However, more recent work by Ji and Held \cite{ji_closure_2013,ji_ion_2015} includes ion-electron collisions in the local limit and finds that they do have a notable impact on the ion heat conductivity and viscosity. Fusion-relevant plasmas often exhibit nonlocal behavior at sharp interfaces; particles with long mean free paths can stream across these interfaces and transport mass, momentum, and energy nonlocally. Nonlocal electron heat conduction has been studied in previous literature through full kinetic simulations \cite{PhysRevLett.46.243,article_brodrick_testing_vfp_snb,10.1063/5.0086783} and closures \cite{Epperlein91,schurtz_nonlocal_2000,holec_nonlocal_2018}, but nonlocal ion transport has not received the same level of attention. Since local ion transport coefficients are significantly impacted by ion-electron collisions, nonlocal ion transport is also expected to be modified. Nonlocal parallel ion transport including ion-electron collisions has been addressed in subsequent work by Ji, Lee, and Held \cite{ji_ion_2017}; however, this approach is limited to the linear response regime in a near-homogeneous plasma and does not extend to other components of magnetized transport. This work applies a \ac{RKM} \cite{Mitchell_2024,mitchell_nonlocal_2025,mitchell_first-principles_2026} to demonstrate the impact of ion-electron collisions on ion transport in the nonlocal regime. 

%An accurate understanding and modeling of nonlocal transport is then essential to the design and improvement of confined fusion schemes.
%While transport in highly collisional plasmas has been studied extensively in previous literature,

Plasmas often contain multiple ion species, giving rise to new transport physics compared to the simple plasma case, with direct implications for the performance of \ac{ICF} experiments \cite{PhysRevLett.105.115005,Kagan_2014,10.1063/1.4745869}. General formalisms for multi-species transport have been established in the local, collisional limit \cite{Ferziger_book,zhdanov_transport_2014}, with compact explicit expressions for multi-species transport coefficients provided by Kagan and Baalrud \cite{kagan2016transport}. Although ion-electron collisions can be readily accounted for in these frameworks, they have not been explicitly studied in works presenting or applying transport coefficients for ionic mixtures as far as the authors are aware. This work explicitly presents the impact of ion-electron collisions on multi-species transport fluxes in both the local and nonlocal regimes.

The remainder of the paper is organized as follows: Sec. \ref{single_species_background} provides an overview of single-species ion transport, emphasizing the significance of ion-electron collisions. The \ac{RKM} for a single ion species is presented in Subsection \ref{rkm_single_subsection}. Results for the impact of ion-electron collisions on nonlocal ion heat conduction and ion viscosity are then presented in Subsections \ref{nonlocal_q_subsection} and \ref{nonlocal_pi_subsection}, respectively. Sec. \ref{sec:multi-species} extends the discussion to ionic mixtures. The \ac{RKM} for mixtures is presented in Subsection \ref{rkm_mixture_subsection} and results for the impact of ion-electron collisions on transport in mixtures are presented in Subsection \ref{mixture_results_subsection}. Sec. \ref{discussion} discusses these results.

\section{Single-species ion transport}\label{single_species_background}

%\subsection{Local transport}

In the strongly collisional limit, local analytic transport closures derived from the Chapman-Enskog expansion are valid. Collisionality is often parametrized by the Knudsen number $N_K = \lambda_{\text{th}}/L$, where $\lambda_{\text{th}}$ is a characteristic thermal \ac{MFP} and $L$ is a characteristic hydrodynamic length scale, often $L\sim \vert \boldsymbol{\nabla} \ln T \vert^{-1}$; local analytic closures are then valid where $N_K\ll 1$.

The local analytic results for the ion heat flux and viscous stress tensor in a simple magnetized plasma have the form
\begin{align}
    \boldsymbol{q}_i^{\text{local}} &= - \kappa_c^i \Big( [\underbrace{\hat{\kappa}_\parallel^i \boldsymbol{b}\boldsymbol{b} + \hat{\kappa}_\perp^i(\mathbb{I}-\boldsymbol{b}\boldsymbol{b})]\cdot\boldsymbol{\nabla}T_i }_{\text{Heat conduction}}  \underbrace{-\hat{\kappa}_\times^i \boldsymbol{b}\times \boldsymbol{\nabla}T_i}_{\text{Righi-Leduc}} \Big),\notag\\[0.5em]
     \Pi_i^\text{local} &= - \eta_c^i \Big( \underbrace{\hat{\eta}_0^i \mathrm{W}_0^i+ \hat{\eta}_1^i \mathrm{W}_1^i  +  \hat{\eta}_2^i \mathrm{W}_2^i}_{\text{Ordinary viscous stress}} \underbrace{ -\hat{\eta}_3^i \mathrm{W}_3^i  - \hat{\eta}_4^i \mathrm{W}_4^i}_{\text{Gyroviscous stress}} \Big),
\end{align}
where $\kappa_c^i = \frac{n_i T_i \tau_{ii}}{m_i}$ and $\eta_c^i = n_i T_i \tau_{ii}$ are the basic ion heat conductivity and viscosity, respectively, with the ion-ion collision time $\tau_{ii} = \frac{3\sqrt{\pi}}{4\hat{\nu}_{ii}} = \frac{6\sqrt{2}\pi^{3/2}\varepsilon_0^2 m_i^{1/2} T_i^{3/2}}{n_i Z^4 e^4 \ln \Lambda_{ii} }$ and Coulomb logarithm $\ln\Lambda_{ii}$. $\hat{\kappa}_{(\parallel,\perp,\times)}^i$ and $\hat{\eta}^i_{(0,\dots,4)}$ are the dimensionless heat conductivity and viscosity coefficients. The tensors $\mathrm{W}_{(0,\dots,4)}^i$ are suitable contractions of $\boldsymbol{b}=\boldsymbol{B}/\vert \boldsymbol{B}\vert$ with the ion rate-of-strain tensor $\mathrm{W}^i = \boldsymbol{\nabla}\boldsymbol{u}_i + (\boldsymbol{\nabla}\boldsymbol{u}_i)^{\text{T}} - \frac{2}{3}(\boldsymbol{\nabla}\cdot\boldsymbol{u}_i)\mathbb{I}$ as defined by Braginskii. Braginskii's viscous stress closure is valid for order-unity Mach number $ M=\vert \boldsymbol{u}_i\vert/v_{\text{th},i} \sim \mathcal{O}(1)$, which this work is limited to \cite{mikhailovskii_transport_1971}.

Transport becomes nonlocal where the plasma is no longer sufficiently collisional. Transport fluxes such as heat flow are predominantly carried by suprathermal particles; since the \ac{MFP} scales strongly with particle speed $\lambda\sim v^4$, nonlocal effects become significant for Knudsen numbers $N_K \gtrsim 1/100$. Nonlocal transport has mostly been studied in the context of electron heat conduction \cite{PhysRevLett.46.243,Epperlein91,schurtz_nonlocal_2000,del_sorbo_reduced_2015}. Heat flow is the most sensitive transport quantity to nonlocal effects since it is carried by higher energy particles, and electrons are far more effective at transporting heat than the much heavier ions. However, there are various scenarios in which the ion heat flux can become comparable to or dominate the electron heat flux. 

Assuming that the ion and electron temperature gradients are collinear, the ratio of the local ion and electron conductive heat fluxes is
\begin{equation}
    \frac{\vert \boldsymbol{q}_{i}^T \vert}{\vert \boldsymbol{q}_{e}^T \vert} = \frac{\hat{\kappa}^i_{(\parallel,\perp,\times)}}{\hat{\kappa}^e_{(\parallel,\perp,\times)}} \bigg( \frac{m_e}{m_i} \bigg)^{1/2} \bigg( \frac{T_i}{T_e} \bigg)^{7/2} \frac{1}{Z^3} \frac{\vert \boldsymbol{\nabla} \ln T_i \vert}{\vert \boldsymbol{\nabla} \ln T_e\vert},
\end{equation}
with the perpendicular ($\perp$) and cross ($\times$) ion and electron dimensionless conductivity coefficients dependent on their respective Hall parameters. Ion heat flow is significant in comparison to electron heat flow where this ratio is comparable to or exceeds unity. The presence of the square root of mass ratio factor $(m_e/m_i)^{1/2}$ means that, taking all other terms as order unity, the ion heat flow is a factor of $ \lessapprox 1/40$ smaller than electron heat flow unless other mechanisms allow it to become comparable or dominant.

In the direction parallel to the magnetic field, the heat flux is independent of magnetization; however, the heat flow in the perpendicular direction is suppressed with magnetic fields. For increasing magnetic field strength, the electron heat flow is magnetized and suppressed significantly more than ion heat flow since electrons are lighter than ions. For large electron Hall parameter $\chi_e \gg 1$, magnetic fields suppress electron heat flow by a factor of $\hat{\kappa}^e_\perp \sim \chi_e^{-2}$ or $\hat{\kappa}^e_\times \sim \chi_e^{-1}$, allowing ion heat flux to dominate electron heat flux. In addition, each species' heat conductivity scales strongly with species temperature; if ions are significantly hotter than electrons, their heat flow can dominate. The strong dependence on the temperature ratio $ (T_i/T_e)^{7/2}$ means even moderate temperature ratios can also cause ion heat transport to become comparable to electron transport, even in the unmagnetized case. Finally, ion temperature gradients can be significantly sharper than electron temperature gradients, for example in the structure of a collisional shock, where the gradient term again can cause the ion heat flow to become comparable to electron heat flow. Taking the ratio of the viscous stress components shows that ion viscous effects typically dominate in low-$Z$ plasmas, with electron viscosity becoming significant in high-$Z$ plasmas \cite{velikovich_role_2001}.

In Braginskii's calculation, where ion-electron collisions are neglected, the resulting magnetized coefficients are dependent only on the ion Hall parameter $\chi_i=Ze \vert \boldsymbol{B}\vert \tau_{ii} / m_i$ and unmagnetized coefficients $(\hat{\kappa}_\parallel^i)^{\text{Braginskii}} /\sqrt{2} = 3.906 $ and $(\hat{\eta}_0^i)^{\text{Braginskii}} /\sqrt{2} = 0.96 $ are numerical constants. However, while the ion-electron collision operator is small by a factor of $\sqrt{m_e/m_i}$, there are other considerations that make ion-electron collisions significant enough to notably modify ion transport coefficients. Consider an ion with normalized speed $x_i=w/v_{\text{th},i}$, where $\boldsymbol{w}=\boldsymbol{v}-\boldsymbol{u}_i$ is the particle velocity in the frame comoving with the local ion Maxwellian bulk velocity, $w=\vert\boldsymbol{w}\vert$ is the particle speed in this frame, and $v_{\text{th},i} = \sqrt{2T_i/m_i}$ is the ion thermal velocity. Ions with speeds in the range $1 \lessapprox x_i \lessapprox 4$ are those that contribute significantly to the heat flux, with slower ions not carrying enough kinetic energy and faster ions being too scarce. Mathematically, this is manifested through the integrand of the local heat flux $dq^i \propto (x_i^2 - 5/2)^2 x_i^3 e^{-x_i^2} dx_i $, which has most of its area over this interval. The speeds of these heat-flow-contributing ions are still much smaller than the electron thermal velocity $w \lessapprox 4 v_{\text{th},i} \ll v_{\text{th},e}$, and therefore there are many more electrons than ions in the nearby velocity space of these suprathermal ions. For a given relative velocity, a single ion-electron collision has a far smaller effect on an ion's behavior than an ion-ion collision since electrons are much lighter than ions. However, Coulomb collisions are more significant for smaller relative velocities between the two colliding particles; ion-electron collisions can therefore become just as significant as ion-ion collisions for suprathermal ions since there are many more electrons than ions in the nearby velocity space. As a result, ion-electron collisions impact ion heat flow. This impact increases with the ion-electron temperature ratio $T_i/T_e$ since electrons become, on average, closer in velocity space to the heat-flow-contributing ions.

%the relative population of thermal electrons to heat-flow-contributing ions increases.

%Since 

%phrase more carefully - expand

%background Maxwellian ions, which have a local phase space density $ \propto n_i m_i^{3/2} T_i^{-3/2} e^{-x_i^2}$ in the nearby velocity space. In contrast, the much lighter electrons have a broader Maxwellian, so these ions collide against electrons with a phase space density $\propto n_e m_e^{3/2} T_e^{-3/2}$. The ratio of between these electron and ion phase space densities is then $\frac{1}{Z} (\frac{m_e}{m_i})^{3/2} (\frac{T_i}{T_e})^{3/2} e^{x_i^2} $. 

\begin{figure}[htpb!]
    \centering    
    \includegraphics[width=0.5\textwidth]{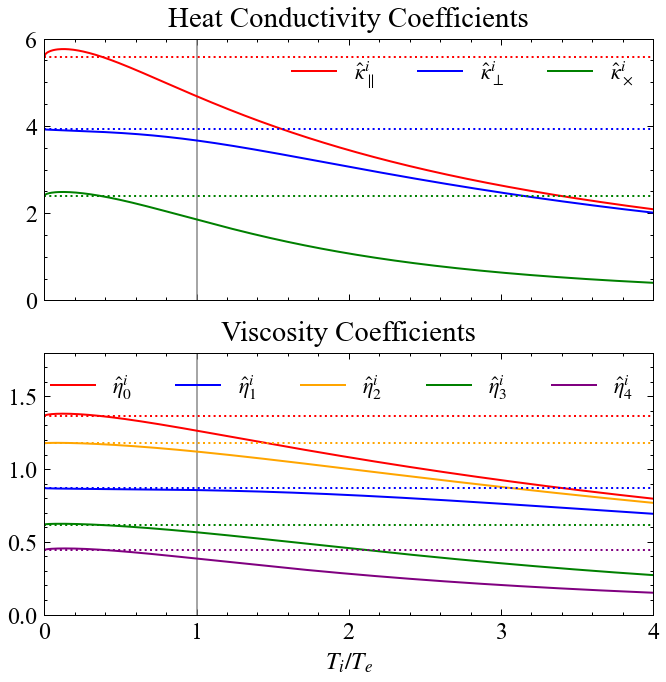}
    \caption{
        Dependence of the dimensionless local ion heat conductivity and viscosity coefficients on ion-electron temperature ratio $T_i/T_e$ for an intermediate magnetization $\chi_i=0.25$, produced using the fits of Ji and Held \cite{ji_ion_2015}. Dotted lines correspond to Braginskii's results where ion-electron collisions are neglected, and therefore represent the limiting values for $T_i/T_e\rightarrow0$. The red lines also correspond to results for unmagnetized ions since the coefficients $\hat{\kappa}_\parallel^i$ and $\hat{\eta}_0^i$ are independent of the Hall parameter.
    }
    \label{fig:temperature_dependence_conductivities}
\end{figure}

Ji and Held \cite{ji_ion_2015} therefore included the ion-electron collision term in a Braginskii-like local calculation, resulting in a dependence of the ion transport coefficients on $\varphi = \frac{1}{Z}\sqrt{m_e/m_i}$ and $\xi = \sqrt{T_i/T_e}$ as well as the ion Hall parameter $\chi_i$. The dependence of the ion transport coefficients on the temperature ratio $\xi$ for a moderately magnetized hydrogen plasma is shown in Fig. \hyperref[fig:temperature_dependence_conductivities]{1}. Even for equal temperatures ($T_i/T_e=1$) in a hydrogen plasma, the unmagnetized coefficients for heat conductivity $\hat{\kappa}^i_\parallel$ and viscosity $\hat{\eta}^i_0$ are reduced by $16\%$ and $7\%$, respectively, when ion-electron collisions are included. For larger ion-electron temperature ratios, these reductions are much more dramatic, for instance, at $T_i/T_e=10$, the transport coefficients $\hat{\kappa}^i_\parallel$ and $\hat{\eta}^i_0$ are reduced by $82\%$ and $71\%$, respectively. Since the ratio of the ion-ion and ion-electron collision frequencies scales linearly with ionization $\hat{\nu}_{ii}/\hat{\nu}_{ie} \propto Z$, the impact of ion-electron collisions becomes weaker in higher-$Z$ plasmas. In confined fusion schemes, various mechanisms such as collisional shocks \cite{zeldovich_physics_1967}, neutral beam injection \cite{koch_plasma_2004}, and plasma waves \cite{hwang_heating_1983} preferentially heat ions over electrons, leading to temperature ratios exceeding unity $T_i/T_e>1$ and making ion-electron collisions even more significant. This work investigates the impact of ion-electron collisions on transport in the nonlocal regime.

\subsection{Reduced kinetic method for a single ion species}\label{rkm_single_subsection}

A single ion species in a simple plasma is described by the \ac{VFP} equation
\begin{equation}
\begin{aligned}\label{ion_VFP}
    \partial_t f_i &+ \boldsymbol{v}\cdot \boldsymbol{\nabla} f_i + \frac{Ze}{m_i}\big(\boldsymbol{E} + \boldsymbol{v}\times\boldsymbol{B}\big)\cdot\boldsymbol{\nabla}_{\boldsymbol{v}}f_i = C_{ii} + C_{ie},
\end{aligned}
\end{equation}
where $C_{ii}$ and $C_{ie}$ are the Fokker-Planck collision operators representing ion-ion and ion-electron collisions, respectively. Previous work \cite{Mitchell_2024,mitchell_nonlocal_2025,mitchell_first-principles_2026} presented an \ac{RKM} which derives and solves a linear equation,
\begin{align}
    D_i = (\hat{C}_{ii} + \hat{C}_{ie} - \hat{V}_i - \hat{B}_i) \delta f_i,
\end{align}
for the deviation of the \ac{DF} from Maxwellian $\delta f_i = f_i-f_i^M$ from which transport fluxes are evaluated. The definition of these terms and details of the \ac{RKM} are provided in \cite{mitchell_first-principles_2026}. The \ac{RKM} is derived directly from the \ac{VFP} equation with the first-principles Fokker-Planck collision operator for small-angle Coulomb collisions. The key assumption of this approach is neglecting the nonlinear component of the collision operator $C_{ii}\{\delta f_i,\delta f_i\}$, which is justified \textit{a posteriori} for Knudsen numbers $N_K \lesssim 1$. Ion-electron collisions are straightforwardly included in the \ac{RKM} through the collision term $\hat{C}_{ie}$. The remainder of this section applies the single-species \ac{RKM} to investigate the impact of ion-electron collisions on the nonlocal ion heat flux and viscous stress tensor while comparing against the local analytic results of Ji and Held \cite{ji_ion_2015}.

%\begin{equation}
%    \mathcal{D}_i^{(l,m)}  = \mathcal{C}_{ii}^{(l,m)} + \mathcal{C}_{ie}^{(l,m)}  - \mathcal{V}_{i}^{(l,m)} - \mathcal{B}_{i}^{(l,m)}.
%\end{equation}

%The ion-electron collision terms are neglected by Braginskii since they are small by a factor of mass-ratio, however Ji and Held have shown that ion-electron collisions have a significant impact ($\sim 10\%$) on ion transport coefficients \cite{10.1063/1.2977983,10.1063/1.4801022}, especially for larger ion temperatures $T_i > T_e$.

\subsection{Conductive heat flux}\label{nonlocal_q_subsection}

\begin{comment}
\begin{figure}[!htb]
  \centering
  \subfloat[Conductive heat flux (unmagnetized)]{%
    \includegraphics[width=0.5\textwidth]{heat_flux_merged.png}%
    \label{fig:non-local_ie}
  }\\
  \subfloat[Conductive heat flux and Righi-Leduc heat flux (magnetized)]{%
    \includegraphics[width=0.5\textwidth]{heat_flux_mag_merged.png}%
    \label{fig:non-local_ie_perp}
  }
  \\
  \subfloat[Ratio of heat fluxes with and without ion-electron collisions]{\includegraphics[width=0.5\textwidth]{heat_flux_ion_ratios_vert.png}
    \label{fig:non-local_heat_flow_ratios}
  }
  \caption{Local ($N_K=0.001$) and nonlocal ($N_K=0.2$) ion heat flux due to a temperature step without ion-electron collisions and with ion-electron collisions for two temperature ratios $T_i/T_e=1,4$. Panel (a) shows the conductive ion heat flux in an unmagnetized hydrogen plasma. Panel (b) shows the perpendicular conductive heat flux and Righi-Leduc heat flux in a moderately magnetized hydrogen plasma with an ion Hall parameter $\chi_{i,0} = 0.25$ at the hot end. Panel (c) shows the ratio of heat fluxes with and without ion-electron collisions for the two temperature ratios. Dotted and solid lines correspond to the local and nonlocal cases respectively. Heat fluxes are appropriately normalized against a characteristic free streaming heat flux $q_0 = n_i T_i^{\text{hot}} v_{\text{th},i}^{\text{hot}}$ and the Knudsen number $N_K$.}
  \label{fig:ion_heat_flux_combined}
\end{figure}
\end{comment}
\begin{figure}[!htb]
  \centering
  \subfloat[Conductive heat flux (unmagnetized)]{%
    \includegraphics[width=0.5\textwidth]{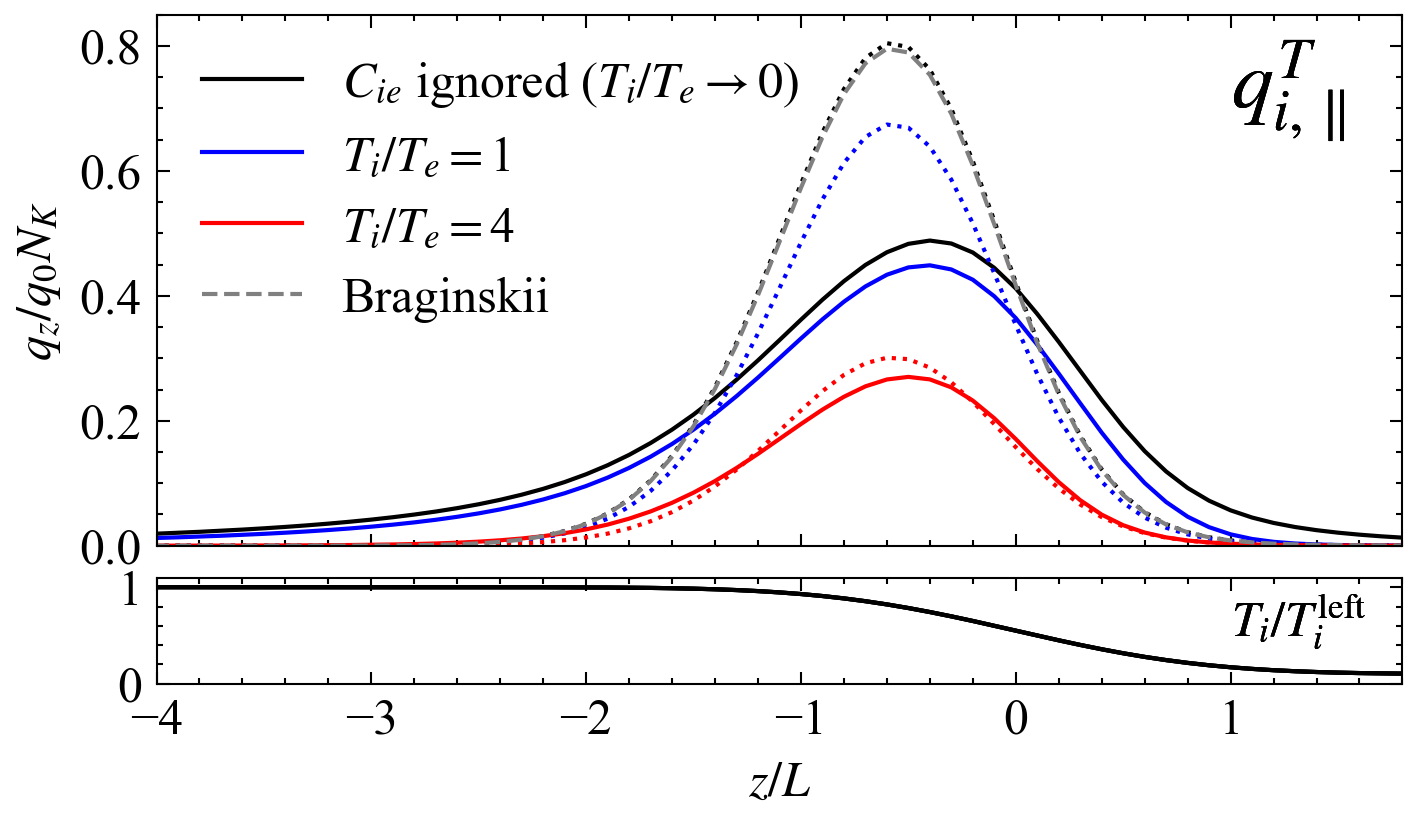}%
    \label{fig:non-local_ie}
  }\\
  \subfloat[Conductive heat flux and Righi-Leduc heat flux (magnetized)]{%
    \includegraphics[width=0.5\textwidth]{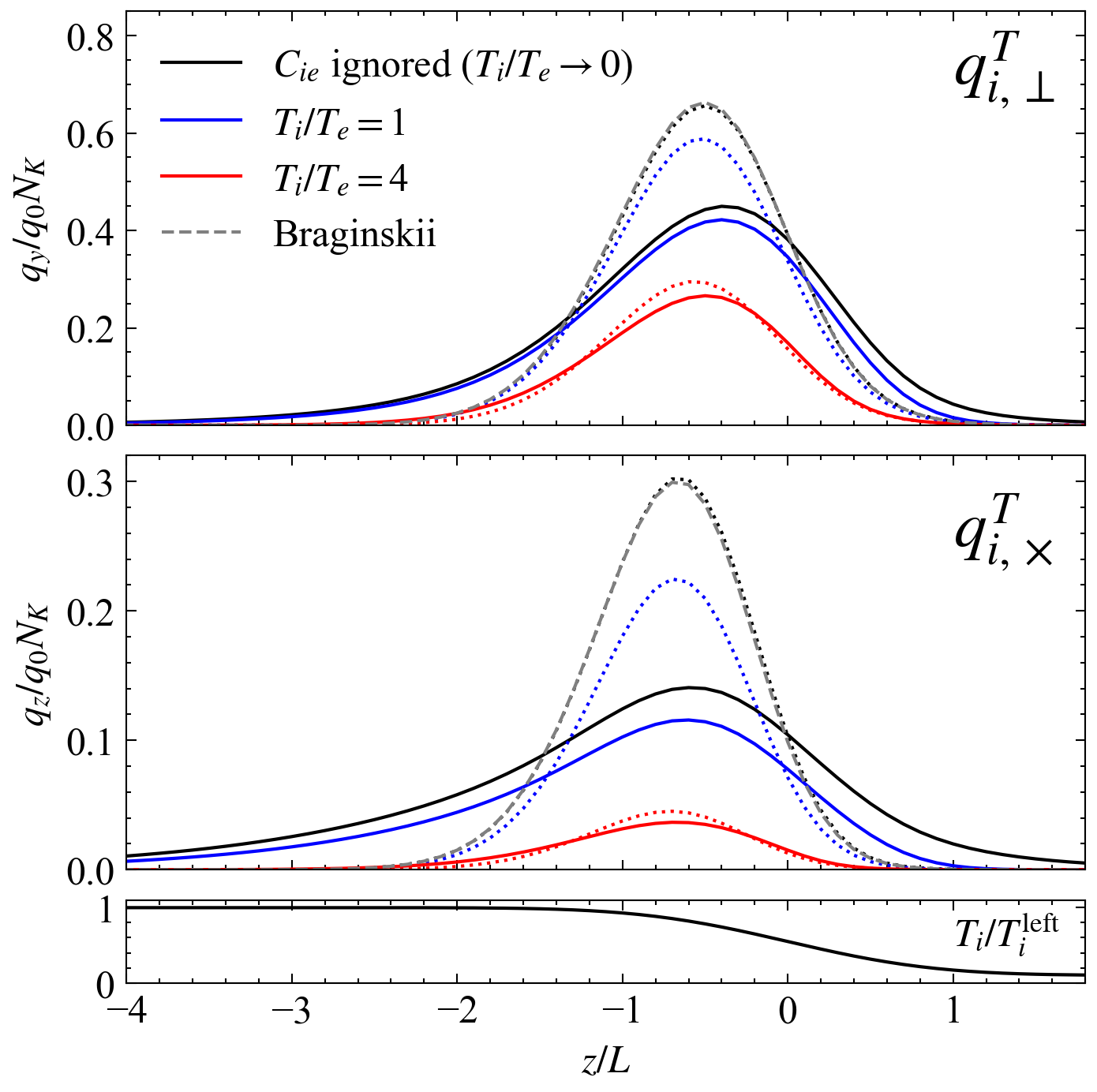}%
    \label{fig:non-local_ie_perp}
  }
  \caption{Local ($N_K\ll 1$, dotted lines) and nonlocal ($N_K=0.2$, solid lines) ion heat flux due to a temperature step without ion-electron collisions and with ion-electron collisions for two temperature ratios $T_i/T_e=1,4$. Panel (a) shows the conductive ion heat flux in an unmagnetized hydrogen plasma. Panel (b) shows the perpendicular conductive heat flux and Righi-Leduc heat flux in a moderately magnetized hydrogen plasma with an ion Hall parameter $\chi_{i,0} = 0.25$ at the hot end. Panel (c) shows the ratio of heat fluxes with and without ion-electron collisions for the two temperature ratios. Heat fluxes are appropriately normalized against a characteristic free streaming heat flux $q_0 = n_i T_i^{\text{hot}} v_{\text{th},i}^{\text{hot}}$ and the Knudsen number $N_K$.}
  \label{fig:ion_heat_flux_combined}
\end{figure}

\addtocounter{figure}{-1}
\begin{figure}[!htb]
  \centering
  \setcounter{subfigure}{2}%
  \subfloat[Ratio of heat fluxes with and without ion-electron collisions]{%
    \includegraphics[width=0.5\textwidth]{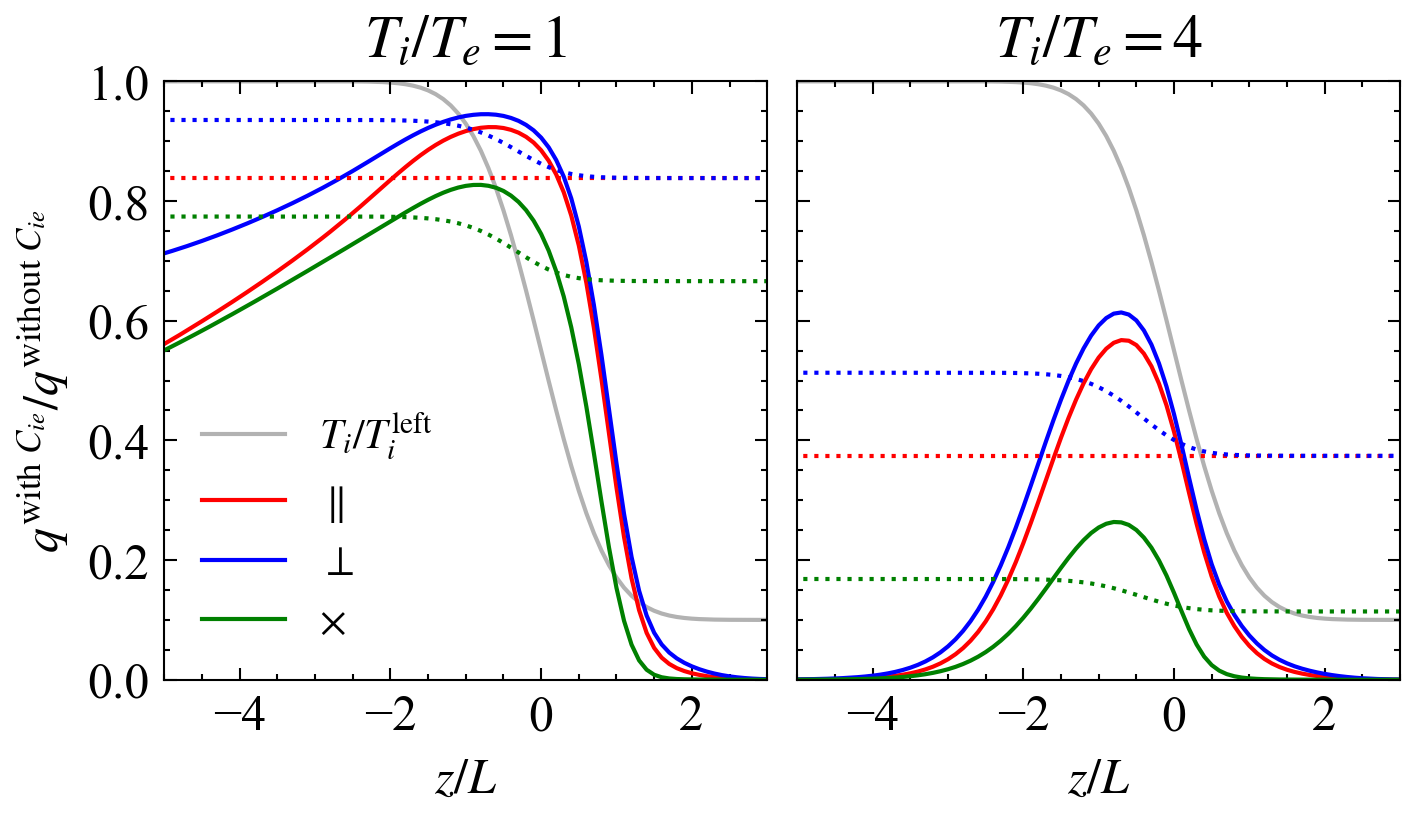}%
    \label{fig:non-local_heat_flow_ratios}
  }
  \caption{(continued)}
\end{figure}

Fig. \hyperref[fig:ion_heat_flux_combined]{2} shows the ion heat flux for an isochoric, simple hydrogen plasma with a temperature profile
\begin{equation}
    \label{temperature_error_function}
    T_i(z)
    = 
    \frac{T_i^{\text{hot}} + T_i^{\text{cold}}}{2} - \frac{T_i^{\text{hot}} - T_i^{\text{cold}}}{2}
    \text{erf}\left(\dfrac{z}{L}\right)
    ,
\end{equation}
where a temperature step $T_i^\text{hot} /T_i^\text{cold} = 10 $ is chosen. The sharpness of the gradient is controlled by the Knudsen number $N_K = \max_z(\lambda_{\text{th},ii}) / L$, where the ion-ion thermal \ac{MFP} is $\lambda_{\text{th},ii} = v_{\text{th},i} / \hat{\nu}_{ii}$. The nonlocal cases are chosen to have a Knudsen number of $N_K=0.2$. Although the electron and ion temperature profiles can be entirely independent, this work considers cases where they are directly proportional across the spatial domain, i.e., $\xi = \sqrt{T_i(z)/T_e(z)}$ is uniform. The result for $T_i/T_e \rightarrow 0$ corresponds to neglecting ion-electron collisions, and therefore matches Braginskii's results and previous work on nonlocal ion heat flux \cite{Mitchell_2024}. Fig. \hyperref[fig:non-local_ie]{2(a)} shows the unmagnetized case, Fig. \hyperref[fig:non-local_ie_perp]{2(b)} shows a moderately magnetized case, and Fig. \hyperref[fig:non-local_heat_flow_ratios]{2(c)} emphasizes the impact by showing the ratio of the different components of heat flow with and without ion-electron collisions for the unmagnetized and magnetized cases.

We first focus on the unmagnetized case in Fig. \hyperref[fig:non-local_ie]{2(a)}. For equal ion and electron temperatures $T_i/T_e = 1$, ion-electron collisions suppress the peak heat flow by $16\%$ in the nonlocal case, similarly to the $16\%$ reduction found by Ji and Held in the local limit. When ion-electron collisions are ignored, there is a significant nonlocal preheat in the cold end for $0.5 \lessapprox z/L \lessapprox 2$. However, this nonlocal preheat is significantly suppressed by ion-electron collisions in the $T_i/T_e \geq 1$ cases. This is explained by the behavior of the slowing collision frequency when ion-electron collisions are included. The ion slowing frequency entering the ion collision operator as $(\hat{C}_i f_i)_{\text{slowing}} = \frac{1}{w^2}\partial_w (w^3 \tilde{\nu}_s^i f_i) $ may be written as
\begin{equation}
    \tilde{\nu}_s^i = \hat{\nu}_{ii} \bigg( \underbrace{\frac{2G(x_i)}{x_i}}_{\text{ion-ion}} + \underbrace{\frac{4}{3\sqrt{\pi}} \frac{\ln\Lambda_{ie}}{\ln{\Lambda}_{ii}} \frac{1}{Z}\sqrt{\frac{m_e}{m_i}} \sqrt{\frac{T_i}{T_e}}}_{\text{ion-electron}}  \bigg),
\end{equation}
where $G(x)$ is the Chandrasekhar function \cite{chandrasekhar_dynamical_1943,helander_and_sigmar,a_beresnyak_2023_2023}. For $x_i \gtrsim 2$, $2G(x_i)/x_i \approx 1/x_i^3$. Suprathermal particles that stream from the hot end and give rise to nonlocal heat flow in the cold end typically have velocities on the order of the hot thermal velocity, say $\sim k v_{\text{th},i}^{\text{hot}}$ where $ 1\lessapprox k \lessapprox4$. In the cold end, these particles have a dimensionless velocity $x_i \sim k 
\sqrt{T_i^{\text{hot}}/T_i^{\text{cold}}} $, i.e., are a factor $\sqrt{T_i^{\text{hot}}/T_i^{\text{cold}}}$ further out into the tail of the ion distribution compared to ions contributing to local heat flow. The ion-ion slowing frequency is small by a factor $\sim \mathcal{O}(1/x_i^{3})$, so ions may stream further into the cold end and give rise to a substantial preheating region. However, when included, the ion-electron slowing frequency does not decay further into the tail. Ion-electron collisions therefore assist and even dominate ion-ion collisions in the slowing and scattering of suprathermal ions. The origin of these particles from the hot end means that the relative importance of ion-electron collisions to ion-ion collisions is enhanced by a factor of $\sim (T_i^{\text{hot}}/T_i^{\text{cold}})^{3/2}$, explaining why the nonlocal preheat is more suppressed by ion-electron collisions than the local heat flux. 

For the larger temperature ratio $T_i/T_e =4$, ion-electron collisions suppress the local result for the heat flux significantly, by $ 63\%$. Increased collisionality from ion-electron collisions then mostly inhibits nonlocal behavior for $N_K = 0.2$, restoring the heat flux to nearly its local result. This is partly due to the definition of the Knudsen number $N_K = \lambda_{\text{th},ii}/L$ used here which does not account for ion-electron effects. %For this temperature ratio, the result for a larger Knudsen number $N_K=0.5$ is therefore also included to show that the heat flux may still be nonlocal for high ion-electron temperature ratios with similar qualitative behavior.

The influence of ion-electron collisions on the magnetized ion heat flow is shown in Fig. \hyperref[fig:non-local_ie_perp]{2(b)}. In these cases, a magnetic field $\boldsymbol{B}=B_0\boldsymbol{\hat{x}}$ is included, giving an ion Hall parameter $\chi_{i,0} = 0.25$ throughout the plasma. The magnetic field suppresses the perpendicular component of the conductive heat flux. The impact of ion-electron collisions on $q^T_{i,\perp}$ is less than that on the unmagnetized heat flow $q^T_{i,\parallel}$ since the suprathermal particles more affected by ion-electron collisions are also more magnetized. Similarly to the unmagnetized ion conductive heat flow, accounting for ion-electron collisions for equal ion and electron temperatures significantly modifies the local result and largely suppresses the preheat for the nonlocal heat flow. For the larger temperature ratio $T_i/T_e=4$, ion-electron collisions reduce the local result of the Righi-Leduc component by $ 70\%$ and $ 90\%$ in the hot and cold ends, respectively. This suppression is stronger than in the unmagnetized case, since the Righi-Leduc heat flow is sourced from particles with larger velocities $w$ than the parallel and perpendicular counterparts. Mathematically, this is apparent from the integrand of the Righi-Leduc heat flux $dq_\times^i/dw$ being weighted to higher $x_i$ than the parallel and perpendicular flux integrands $d q_{(\parallel,\perp)}^i/dw$.

\subsection{Viscous stress tensor}\label{nonlocal_pi_subsection}

The ion viscous stress tensor is also influenced by ion-electron collisions and nonlocal effects. To investigate the impact of these two effects, a simple hydrogen plasma with uniform density and temperature is considered with a shear flow profile
%\begin{align}
%\boldsymbol{u}_i(z) = u_0 \frac{ 3 - \,\text{erf}%(z/L) }{4}\boldsymbol{\hat{y}},   
%\end{align}
\begin{align}
    \boldsymbol{u}_i(z)= \bigg(
    \frac{u^{\text{left}} + u^{\text{right}}}{2} - \frac{u^{\text{left}} - u^{\text{right}}}{2}
    \text{erf}\left(\dfrac{z}{L}\right)\bigg) \boldsymbol{\hat{y}},
\end{align}
with $ u^{\text{left}} =2 v_{\text{th},i} $ and $u^{\text{right}}=v_{\text{th},i}$. Throughout this profile, the Mach number is of order unity $1\leq M\leq 2$, for which Braginskii's result for the ion viscous stress is valid.

\begin{figure}[!htb]
  \centering
  \subfloat[Ordinary viscous stress component (unmagnetized)]{%
    \includegraphics[width=0.5\textwidth]{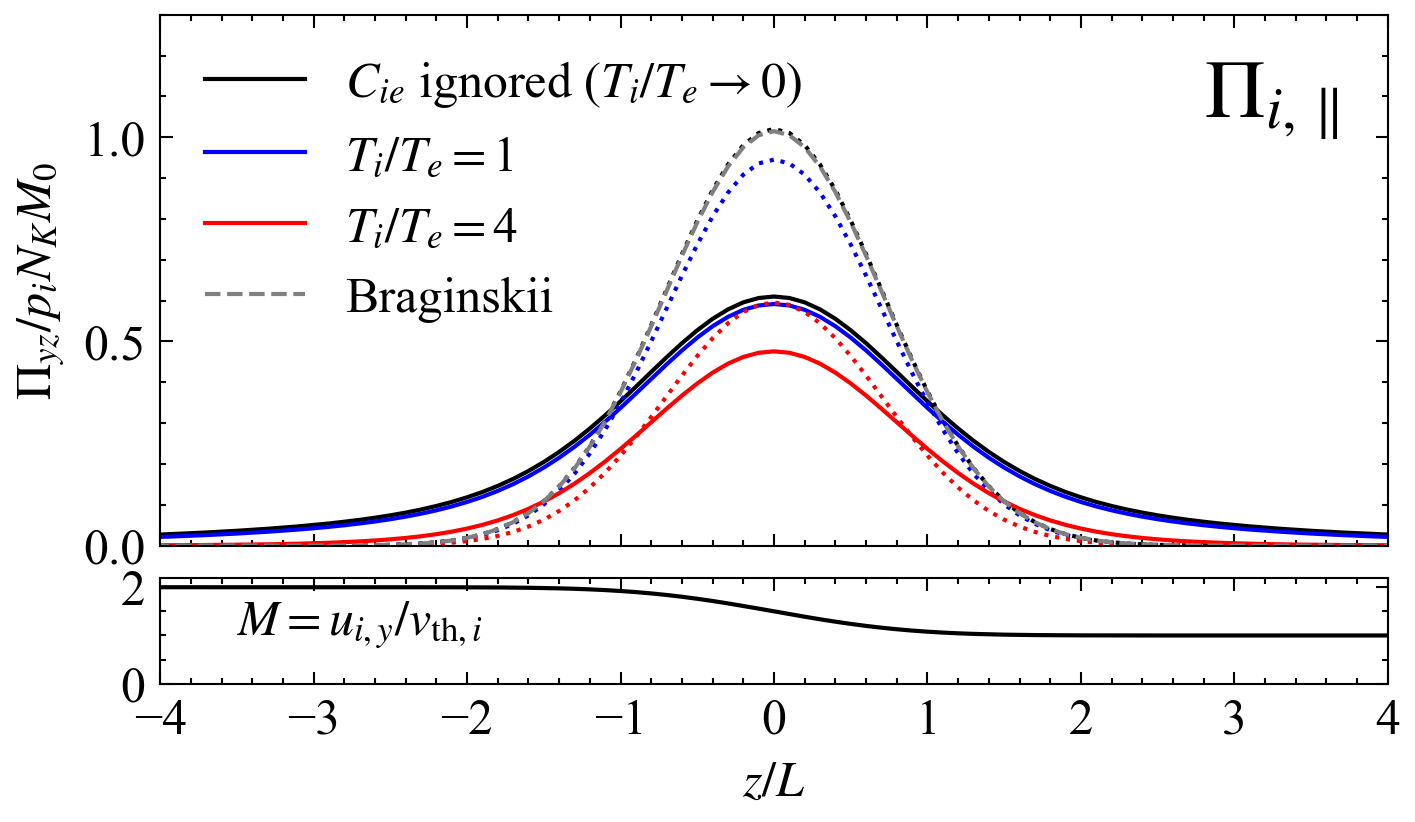}%
    \label{fig:non-local_viscosity}
  }\\
  \subfloat[Ordinary viscous stress and gyroviscous stress components (magnetized)]{%
    \includegraphics[width=0.5\textwidth]{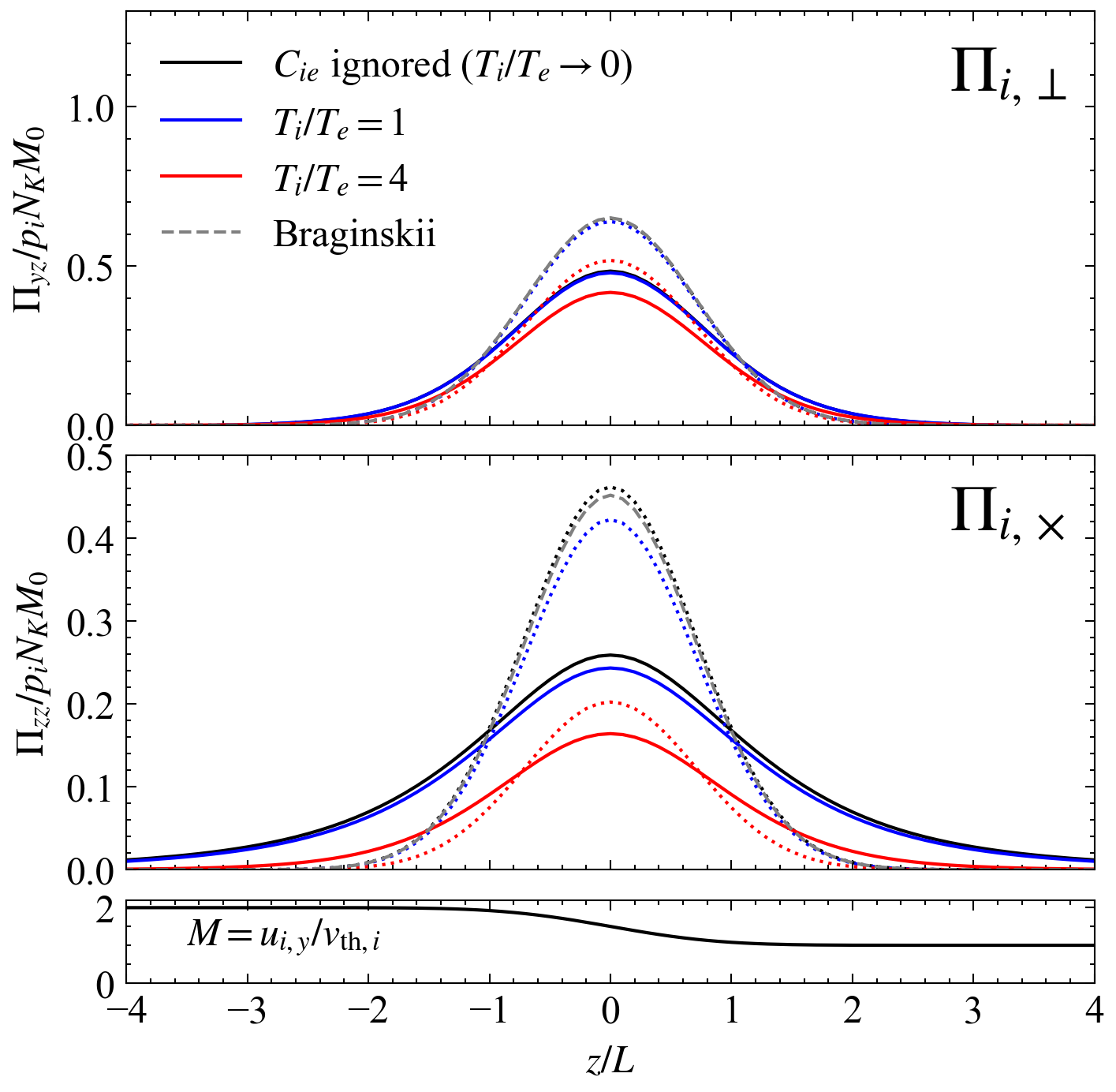}%
    \label{fig:non-local_viscosity_magnetised}
  }
  \caption{Local ($N_K\ll 1$, dotted lines) and nonlocal ($N_K=0.5$, solid lines) ion viscous stress due to a shear flow without ion-electron collisions and with ion-electron collisions for two temperature ratios $T_i/T_e=1,4$. Panel (a) shows the viscous stress $\Pi_{yz}$ in an unmagnetized hydrogen plasma. Panel (b) shows the ordinary viscous stress component $\Pi_{i,yz}$ and gyroviscous stress component $\Pi_{i,zz}$ in a moderately magnetized hydrogen plasma with ion Hall parameter $\chi_i=0.25$. The viscous stress components are appropriately normalized by the ion pressure $p_i=n_iT_i$, Knudsen number $N_K$, and peak Mach number $M_0=u^{\text{left}}/v_{\text{th},i}=2$. The ion flow profile $u_{i,y}(z)$ is shown in the bottom row of subplots.}
  \label{fig:non-local_viscosity_combined}
\end{figure}
%\ref{fig:non-local_viscosity_combined}
Fig. \hyperref[fig:non-local_viscosity_combined]{3} shows the viscous stress tensor $\Pi_{i}$ in a simple hydrogen plasma for different ion-electron temperature ratios, with the unmagnetized and magnetized cases shown in Figs.  \hyperref[fig:non-local_viscosity]{3(a)} and \hyperref[fig:non-local_viscosity_magnetised]{3(b)}, respectively. The behavior is qualitatively similar to the previous heat flow cases; when ion-electron collisions are neglected, the results of Ji-Held and Braginskii agree well, but for temperature ratios of unity and higher, the stress is suppressed by ion-electron collisions. Nonlocal effects for a high Knudsen number $N_K=0.5$ suppress the peak stress further, but also give rise to an enhanced stress away from the main gradient at $z=0$, analogously to preheat effects in nonlocal heat flow. This enhanced viscous stress increases momentum transfer away from the main region of shearing flow. For equal temperatures, this preheat-like enhanced stress is not noticeably modified, but is strongly suppressed at higher ion-electron temperature ratios in the $T_i/T_e=4$ case where ion-electron collisions almost entirely restore the viscous stress to the local result even for the high Knudsen number $N_K=0.5$. This is partly due to our definition of Knudsen number, which does not include ion-electron collisions but has been chosen to allow a straightforward comparison between different cases.
%\ref{fig:non-local_viscosity_magnetised}

Fig. \hyperref[fig:non-local_viscosity_magnetised]{3(b)} shows the ordinary viscous stress component $\Pi_{i,yz}$ and gyroviscous component $\Pi_{i,zz}$ for the shear flow with a magnetic field $\boldsymbol{B}=-B_0\boldsymbol{\hat{x}}$, with $B_0$ chosen such that the Hall parameter across the domain is $\chi_i = 0.25$. Similarly to the magnetized heat flow, the magnetic field inhibits the shear stress $\Pi_{i,yz}$, with the impact of ion-electron collisions diminished. The gyroviscous stress appears to have similar sensitivity to both nonlocal effects and ion-electron collisions as the unmagnetized viscous stress, differing largely from the standard Braginskii result due to high Knudsen number $N_K = 0.5$ or ion-electron temperature ratio $T_i/T_e=4$.

\section{Multi-species ion transport}\label{sec:multi-species}
%In both magnetically and inertially confined fusion plasmas
The above studies have been limited to a single ion species. However, confined fusion plasmas typically rely on nuclear reactions between different species such as deuterium and tritium; multi-species effects can then lead to species separation which directly reduces fusion reactivity and total yield. In addition, high-$Z$ species can diffuse or stream into the fusion fuel, increasing radiative losses and degrading yield. Multi-species transport effects are therefore a key consideration in fusion plasmas \cite{PhysRevLett.105.115005,Kagan_2014,10.1063/1.4745869,kagan_influence_of_coupling_2016}. Considering an ionic mixture introduces additional physics compared to the single ion species case. Different species may diffuse relative to each other even if the ionic mixture's other state variables, namely the mass density $\rho_i=\sum_\alpha \rho_\alpha$, hydrodynamic velocity $\boldsymbol{u}_i = \frac{1}{\rho_i} \sum_\alpha \rho_\alpha \boldsymbol{u}_\alpha$, and common temperature $T_\alpha = T_i$, are fixed, which changes the composition and therefore the behavior of the mixture. 

In the following, binary mixtures of a light and heavier species, denoted by `$l$' and `$h$' respectively,  are considered. The composition of a binary mixture can be characterized by the mass concentration of the lighter species $c=\rho_l/\rho_i$ which evolves according to its hydrodynamic equation
\begin{align}
    \partial_t c + \boldsymbol{u}_i\cdot\boldsymbol{\nabla}c + \frac{1}{\rho_i}\boldsymbol{\nabla}\cdot\boldsymbol{i} = 0 ,
\end{align}
where $\boldsymbol{i}=\rho_l (\boldsymbol{u}_l-\boldsymbol{u}_i)$ is the diffusive mass flux. Its local closure can be written in the Landau-Lifshitz form \cite{Landau1987Fluid,Kagan_2014,10.1063/1.4745869,Ferziger_book,PhysRevLett.105.115005}
%\begin{align}
%    \boldsymbol{i}^\text{local} &= \rho_i D (\boldsymbol{\nabla}c + k_T \boldsymbol{\nabla}\ln T_i + k_p\boldsymbol{\nabla}\ln p_i),
%\end{align}
\begin{align}
    \boldsymbol{i}^\text{local} &= -\rho_i D (\underbrace{\boldsymbol{\nabla}c}_{\substack{\text{Classical} \\ \text{diffusion}}}  + \underbrace{k_p \boldsymbol{\nabla}\ln p_i}_{\substack{\text{Baro-} \\ \text{diffusion}}}+ \underbrace{ k_T \boldsymbol{\nabla} \ln T_i}_{\substack{\text{Thermo-} \\ \text{diffusion}}} ),
\end{align}
where $D$ is the classical diffusion coefficient, and $k_T$ and $k_p$ are the thermo- and baro-diffusion ratios, respectively. This form emphasizes that diffusion is not only driven by concentration gradients, but also by temperature and pressure gradients. The heat flux of the ionic mixture may be written in an Onsager-symmetric form
\begin{align}
    \boldsymbol{q}_i^\text{local} &= - T_i\kappa_i (\underbrace{  \boldsymbol{\nabla} \ln T_i}_{\substack{\text{Thermal} \\ \text{conduction}}}+ \underbrace{h_c\boldsymbol{\nabla}c}_{\substack{\text{Dufour} \\ \text{heat flow}}}  + \underbrace{h_p \boldsymbol{\nabla}\ln p_i}_{\substack{\text{Baro-} \\ \text{heat flow}}} ),
\end{align}
where $\kappa_i$ is the heat conductivity of the ionic mixture and the dimensionless coefficients $h_c$ and $h_p$ are introduced in a similar manner to the thermo- and baro-diffusion ratios. Ion heat flow in mixtures is then not only driven by temperature gradients, but also by concentration gradients and pressure gradients. The remainder of this section investigates the impact of ion-electron collisions on local and nonlocal multi-species transport in weakly and moderately asymmetric mixtures. Here, the local multi-species results are obtained from the mixture formalism presented by Ferziger and Kaper with ion-electron collisions included \cite{Ferziger_book,kagan2016transport}.

%As far as the authors are aware, previous literature has not explicitly presented or investigated the impact of ion-electron collisions on multi-species transport even in the local regime.
\subsection{Reduced kinetic method for ionic mixtures}\label{rkm_mixture_subsection}

Following previous work \cite{Mitchell_2024,mitchell_first-principles_2026,PhysRevLett.115.105002}, the \ac{RKM} for unmagnetized ionic mixtures solves the set of linear equations
\begin{align}
    D_\alpha = \Big( \Big(\sum_\beta\hat{C}_{\alpha\beta}^T\Big) + \hat{C}_{\alpha e}^T - \hat{V}_\alpha \Big) \delta f_\alpha +  \sum_\beta\hat{C}_{\alpha\beta}^F \delta f_\beta
\end{align}
where $\alpha,\beta$ are ion species labels, $\hat{V}_\alpha$ is the Vlasov term, and $\hat{C}_{\alpha\beta}^T$ and $\hat{C}_{\alpha\beta}^F$ are the test-particle and field-particle collision operators, respectively. The inhomogeneous driving term is
\begin{align}\label{driving_term}
    D_\alpha \equiv \Big( \frac{n_i}{n_\alpha}\boldsymbol{d}_\alpha + (x_\alpha^2-5/2) \boldsymbol{\nabla}\ln T_i \Big) \cdot \boldsymbol{w} f^M_\alpha,
\end{align}
where $x_\alpha=w/v_{\text{th},\alpha}$ and flow gradient terms have been omitted since the viscous stress tensor has already been investigated in the single species context. The concentration gradients, pressure gradient, and electric field appear in the diffusive driving force $\boldsymbol{d}_\alpha \equiv \frac{\boldsymbol{\nabla} p_\alpha - c_\alpha \boldsymbol{\nabla} p_i}{p_i} - c_\alpha \frac{\rho_i}{p_i}(\boldsymbol{X}_\alpha - \sum_\beta c_\beta \boldsymbol{X}_\beta) $, where $\boldsymbol{X}_\alpha$ is the external body force acting on species $\alpha$. Although ionic mixtures can also be magnetized, only unmagnetized mixtures are considered here.

\subsection{Diffusive mass flux and heat flux in mixtures}\label{mixture_results_subsection}

\begin{comment}
\begin{figure}[h]
  \centering
  \subfloat[Classical diffusion / Dufour heat flow ($\nabla c$)]{%
    \includegraphics[width=0.5\textwidth]{grad_c_Cie_all2.png}%
    \label{fig:grad_c}
  }\\
  \subfloat[Thermo-diffusion / heat conduction ($\nabla T_i$)]{%
    \includegraphics[width=0.5\textwidth]{grad_T_Cie_all2.png}%
    \label{fig:grad_T}
  }\\
  \subfloat[Baro-diffusion / baro-heat flow ($\nabla p_i$)]{%
    \includegraphics[width=0.5\textwidth]{grad_p_Cie_all2.png}%
    \label{fig:grad_p}
  }
  \caption{Diffusive mass flux and mixture heat flux for a DT and DC mixture, in three cases driven by a concentration gradient $\nabla c$, temperature gradient $\nabla T_i$, and pressure gradient $\nabla p_i$. Dotted and solid lines correspond to local ($N_K\rightarrow0$) and nonlocal ($N_K=0.2$) results respectively. Each subplot has an independent vertical scale since the absolute magnitude of the flux is difficult to compare between different cases.}\label{mixture_C_ie}
\end{figure}
\end{comment}

\begin{figure}[h!]
  \centering
  \subfloat[Classical diffusion and Dufour heat flow ($\nabla c$)]{%
    \includegraphics[width=0.5\textwidth]{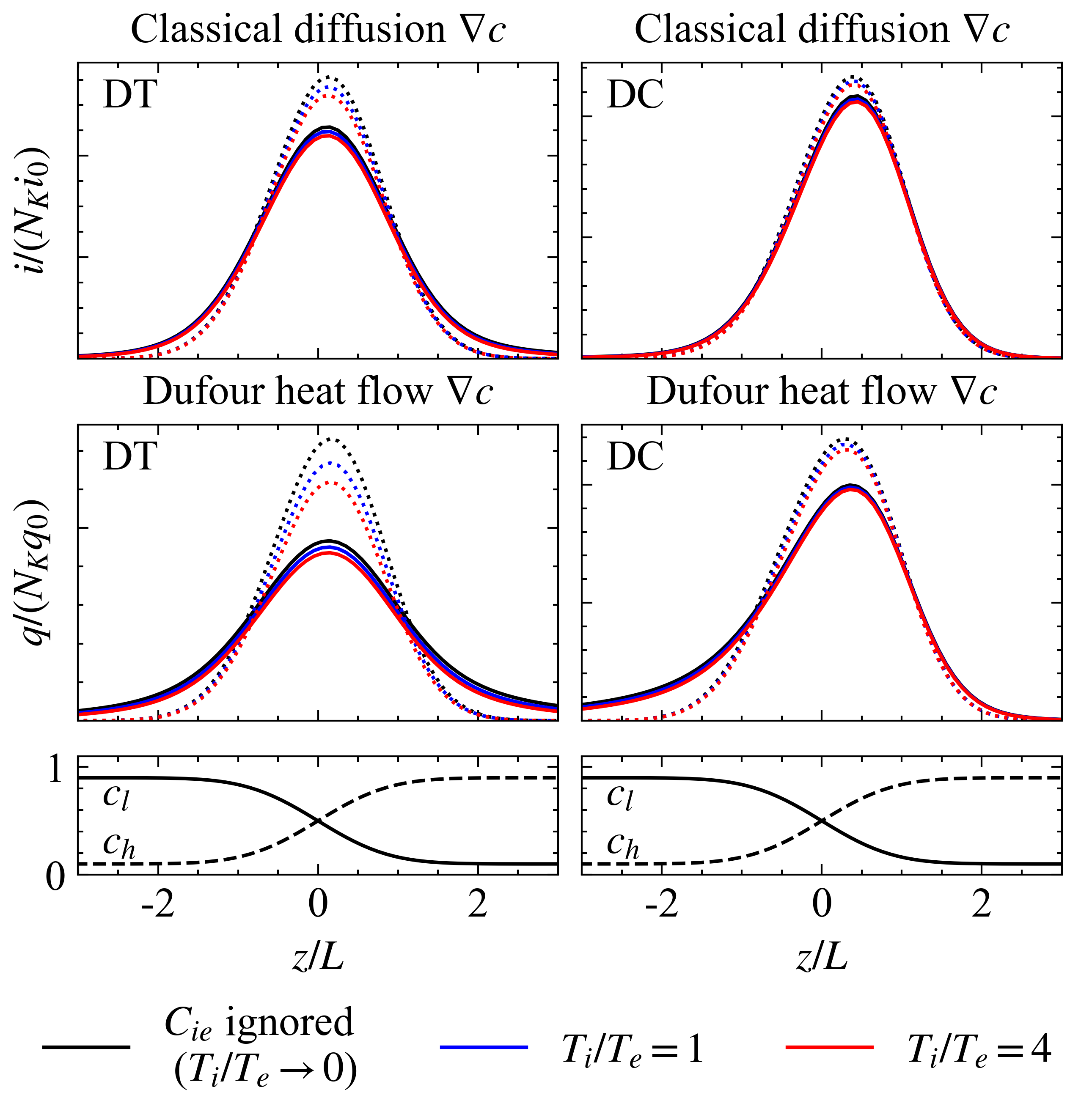}%
    \label{fig:grad_c}
  }\\
  \subfloat[Thermo-diffusion and heat conduction ($\nabla T_i$)]{%
    \includegraphics[width=0.5\textwidth]{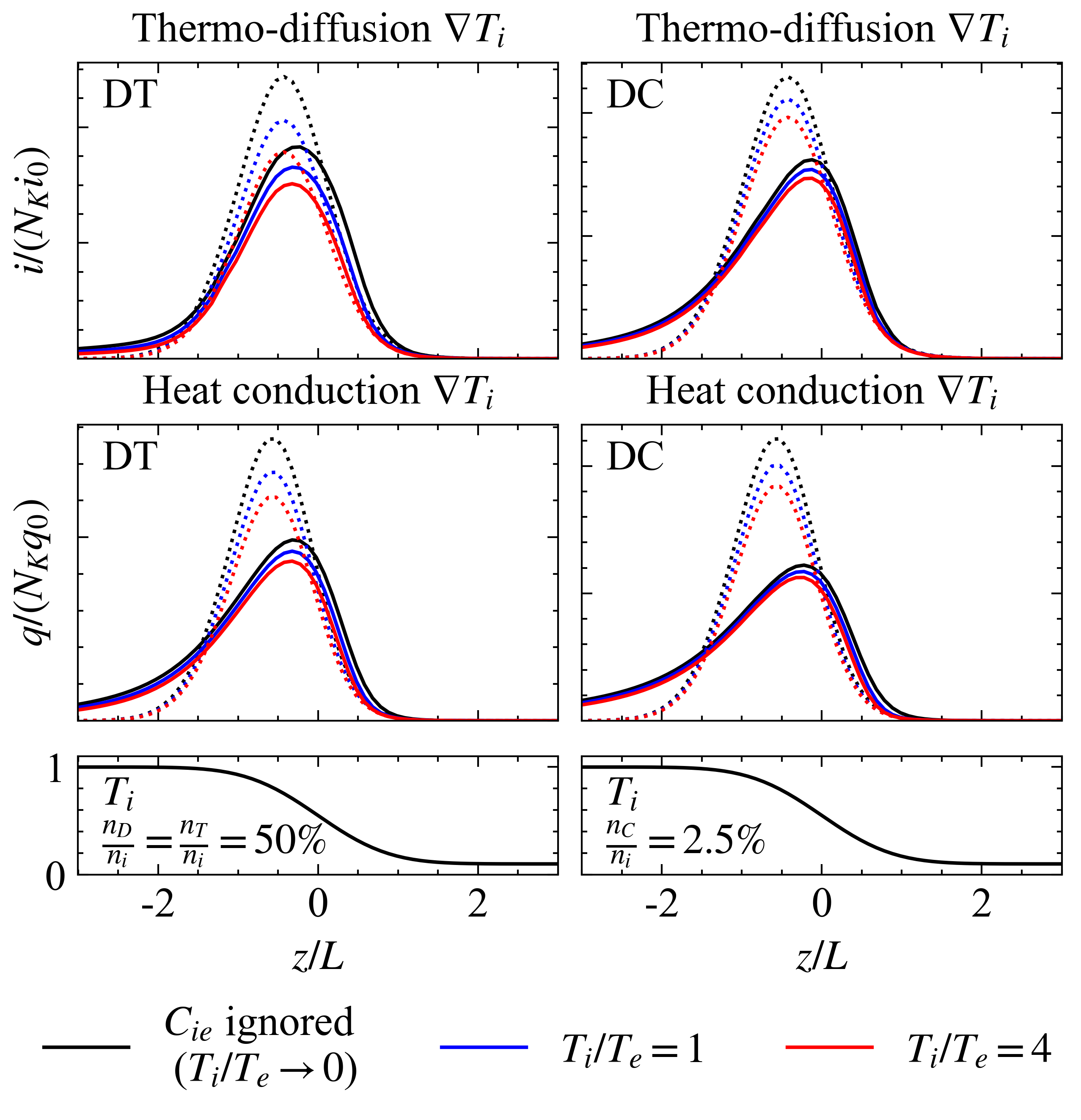}%
    \label{fig:grad_T}
  }
  \caption{Diffusive mass flux and mixture heat flux for a DT and DC mixture, in three cases driven by a concentration gradient $\nabla c$, temperature gradient $\nabla T_i$, and pressure gradient $\nabla p_i$. Dotted and solid lines correspond to local ($N_K\ll 1$) and nonlocal ($N_K=0.2$) results, respectively. Each subplot has an independent vertical scale since the absolute magnitude of the flux is difficult to compare between different cases.}\label{mixture_C_ie}
\end{figure}

\addtocounter{figure}{-1}
\begin{figure}[h]
  \centering
  \setcounter{subfigure}{2}%
  \subfloat[Baro-diffusion and baro-heat flow ($\nabla p_i$)]{%
    \includegraphics[width=0.5\textwidth]{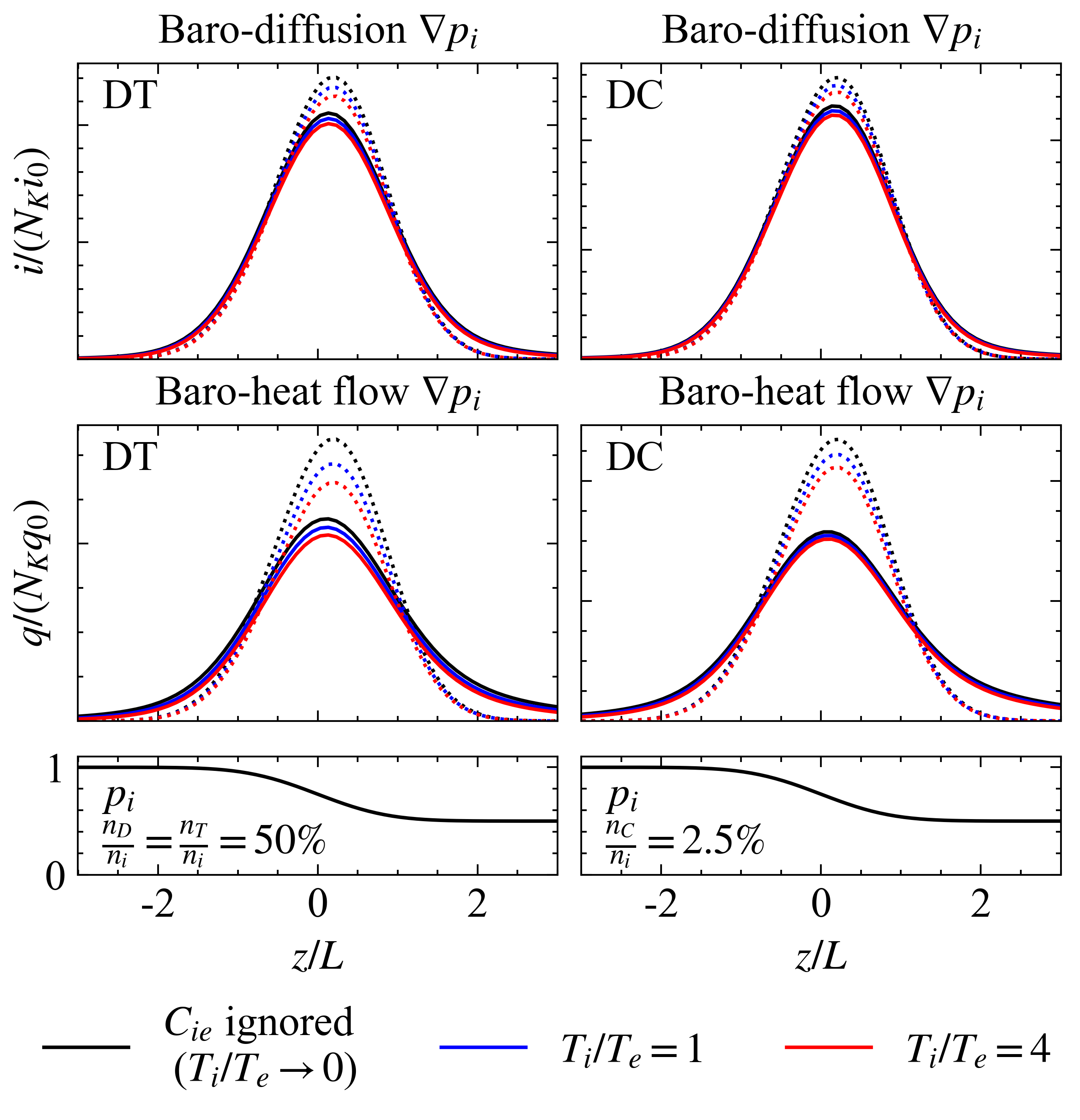}%
    \label{fig:grad_p}
  }
  \caption{(continued)}
\end{figure}
%Fig. \hyperref[mixture_C_ie]{4} shows t
%Equation (\ref{temperature_error_function})
The diffusive mass flux $i$ and heat flux $q_i$ driven by a concentration gradient, a temperature gradient, and a pressure gradient are shown in Figs. \hyperref[fig:grad_c]{4(a)}, \hyperref[fig:grad_T]{4(b)}, and \hyperref[fig:grad_p]{4(c)}, respectively. A weakly asymmetric mixture (DT) and a moderately asymmetric mixture (DC) are considered. All species are assumed to be fully ionized. In each case, the other two profiles are uniform so their thermodynamic drives are zero, i.e., in the $\nabla c$ case the total ion temperature and pressure are uniform, in the $\nabla T_i $ case the concentration and total ion pressure are uniform, and in the $\nabla p_i$ case the concentration and ion temperature are uniform. The profile of the hydrodynamic drive of focus then has an error-function-like profile analogous to Eq. \eqref{temperature_error_function}, with a detailed summary provided in Appendix \ref{appendix_profile_summary}. The electric field is set to zero to ensure consistency between cases. In the DT thermo- and baro-diffusion cases, the deuterium and tritium are in a 50:50 mix by number density, i.e., $n_D /n_i = n_T/n_i = 0.5$. In the DC thermo- and baro-diffusion cases, carbon is chosen to be a $2.5\%$ impurity by number density, i.e., $n_C / n_i = 0.025$, such that the collision frequency against both species $\hat{\nu}_{\alpha\beta}\propto n_\beta Z_\beta^2$ is similar.

Transport in ionic mixtures is dictated by both like-like and like-unlike collisions. To account for the effect of different species, the definition of the Knudsen number is extended to mixtures as
\begin{align}\label{NK_mixture}
    N_K = \max_{z,\alpha} (\lambda_{\text{th},\alpha} )/L, \quad \lambda_{\text{th},\alpha} = \frac{v_{\text{th},\alpha}}{\sum_\beta \hat{\nu}_{\alpha\beta}},
\end{align}
where $\hat{\nu}_{\alpha\beta} = \frac{n_\beta q_\alpha^2 q_\beta^2 \ln\Lambda_{\alpha\beta}}{4\pi\varepsilon_0^2 m_\alpha^2 v_{\text{th},\alpha}^3}$ is the basic collision frequency of species $\alpha$ against species $\beta$. This definition accounts for collisionality against both ion species. In the nonlocal cases, the length scale $L$ of the step is chosen such that $N_K = 0.2$.

In the local case, the impact on classical diffusion and baro-diffusion is identical since the baro-diffusion coefficient $k_p$ is purely thermodynamic, with collisions entering only through the classical diffusion coefficient $D$. Classical diffusion and baro-diffusion are slightly reduced by ion-electron collisions in both mixtures, for both the local and nonlocal cases, whereas thermo-diffusion is significantly affected, with a $ 15\%$ reduction of peak diffusive flux for equal temperatures in the local DT case, rising to $25\%$ for $T_i/T_e=4$. This is expected since the temperature-gradient drive enters the inhomogeneous driving term, given by Eq. \eqref{driving_term}, in the kinetic equation for $\delta f_\alpha$ with an additional $(x_\alpha^2-5/2)$ factor compared to classical diffusion and baro-diffusion. Therefore, temperature gradients drive diffusion of ions in higher velocity regions of the tail where ion-electron collisions are more significant relative to ion-ion collisions. Analogously to nonlocal heat flux giving rise to preheat, nonlocal thermo-diffusion has a preheat-like enhancement of the diffusive flux away from the main gradient region due to streaming suprathermal ions. This `preheat' corresponds to nonlocal transport of impurities, which are known to have a significant impact on both inertially and magnetically confined fusion plasmas. Similarly to the conductive heat flux in the single-species case, this `preheat' is significantly reduced by ion-electron collisions in the cold end of the plasma. 

Since heat flux is a higher velocity moment of the \ac{DF}, it is more sensitive to both nonlocal effects and ion-electron collisions. As before, the temperature gradient term driving the conductive heat flux appears with the same additional $(x_\alpha^2-5/2)$ factor compared to the other thermodynamic drives. Consequently, heat conduction is more sensitive to nonlocality and ion-electron collisions than heat flow driven by other thermodynamic forces, namely Dufour heat flow and baro-heat flow. However, these other heat flows are still noticeably affected by ion-electron collisions in both the local and nonlocal cases, with reductions of peak heat fluxes by $\sim 10\%$. Since the Knudsen number for mixtures defined by Eq. (\ref{NK_mixture}) reduces to the single-species Knudsen number definition for a single species, the nonlocal conductive heat flux in the DT case in Fig. \hyperref[fig:grad_T]{4(b)} is quantitatively similar to the single-species case in Fig. \hyperref[fig:non-local_ie]{2(a)}. Although both species in the ionic mixture transport heat, the lighter species carries a larger proportion of the heat flux. Since a DT mixture is weakly asymmetric, the deuterium and tritium species carry a similar proportion of the heat flux and are similarly influenced by ion-electron collisions. %However, in the DC case, the carbon carries little heat flux compared to the deuterium, and is also practically uninfluenced by ion-electron collisions, unlike the deuterium. As a result, the heat flow is more sensitive to ion-electron collisions in the DT case than in the DC case. 

In ionic mixtures, the ionic heat flux $\boldsymbol{q}_i$ can be decomposed into two components, $\boldsymbol{q}_i=\boldsymbol{q}_i' + \boldsymbol{h}_i$, where 
\begin{align}
    \boldsymbol{q}'_i &= \sum_\alpha \int d^3\boldsymbol{w}\, \bigg(\frac{1}{2}m_\alpha w^2 - \frac{5}{2}T_i\bigg) \boldsymbol{w} f_\alpha
    , \\ \boldsymbol{h}_i &= \sum_\alpha \frac{5}{2}T_i \int d^3\boldsymbol{w}\, \boldsymbol{w} f_\alpha =\sum_\alpha \frac{5}{2}n_\alpha T_i \frac{\boldsymbol{i}_\alpha}{\rho_\alpha}
\end{align}%(\boldsymbol{u}_\alpha-\boldsymbol{u}_i)
are the reduced heat flux and enthalpy flux, respectively. The enthalpy flux is the heat transported due to the inter-diffusion of different species in the ion center-of-momentum frame. Since the enthalpy flux is a lower velocity moment of the \ac{DF} than the reduced heat flux, i.e., it is carried by less energetic particles in the velocity distribution, it is less affected by ion-electron collisions. Ion-electron collisions increase the total collisionality and therefore almost always reduce transport, so the enthalpy flux is less suppressed by ion-electron collisions than the reduced heat flux, increasing its relative significance, especially for larger ion-electron temperature ratios. This can be seen in Fig. \hyperref[mixture_C_ie]{4} since $\boldsymbol{h}_i \propto \boldsymbol{i}_l$ in a binary mixture, and $\boldsymbol{i}_l$ is less sensitive to ion-electron collisions than $\boldsymbol{q}_i$. For example, in the DT $\nabla c$ case in the local limit, ion-electron collisions reduce the enthalpy flux $\boldsymbol{h}_i$ by $5\%$ whereas the reduced heat flux $\boldsymbol{q}_i'$ is reduced by $ 30\%$.

%emphasise similarity to the single species case with this Knudsen number, heat carried by lighter species

%, increasing the overall contribution of the enthalpy flux to the total heat flux from $ 63\%$ to $70\%$.

%ion-electron collisions have weaker impact on enthalpy flux, increasing its significance

%, since ion-electron collisions are more effective at slowing the tritons than the heavier, higher-$Z$ carbon ions.

%Ion-electron collisions still have only a minor effect on the heat flow for equal ion-electron temperatures. The correction in the DT case is noticeable for all three thermodynamic drives, again with a suppression of $  10\%$, increasing to $  20\%$ for $T_i/T_e=4$. In nonlocal cases, the reduction of the heat flux due to ion-electron collisions is similar to the corresponding local results.

%heat flux due to lighter species in moderate asym. mixture

\section{Discussion}\label{discussion}

In summary, the ion heat flux, viscous stress, and diffusive mass flux are significantly influenced by ion-electron collisions in nonlocal cases, similarly to the behavior in the local limit. The \ac{RKM} from previous work captures this effect in a first-principles framework and, to the authors' knowledge, is the only nonlocal closure that models multi-species effects. The nonlocal peak heat flow is suppressed on the order of $16\%$ by ion-electron collisions for equal ion and electron temperatures, rising to $63\%$ for $T_i/T_e=4$, whereas nonlocal preheats are strongly suppressed by ion-electron collisions even for equal ion-electron temperatures. The nonlocal viscous stress is also reduced at its peak and, for larger ion-electron temperature ratios, away from the main flow gradient. Such effects on ion heat transport are particularly relevant to both inertially and magnetically confined fusion plasmas where magnetic fields suppress electron heat flow and high ion temperatures allow ion heat flow to significantly contribute to energy transport, thereby influencing the collective plasma behavior. The impact on the diffusive mass flux is especially significant in confined fusion plasmas, where different species such as deuterium and tritium can separate in space or high-$Z$ impurities can contaminate the fusing plasma, increasing radiative losses and degrading confinement.

\begin{widetext}

\appendix

\section{Summary of Profiles for Mixture Cases}\label{appendix_profile_summary}

For a quantity $Q(z)$, we define an error-function-like profile
\begin{align}
    Q^{\text{erf.}}(z \vert Q_{\text{left}}, Q_{\text{right}},L) =  \frac{Q_{\text{left}} + Q_{\text{right}}}{2} - \frac{Q_{\text{left}} - Q_{\text{right}}}{2}
    \text{erf}\left(\dfrac{z}{L}\right).
\end{align}
The profiles in the multi-species cases of Section \ref{sec:multi-species} are then summarized in Table \ref{table:profiles}.

\FloatBarrier
\begin{table*}[]
{\setlength{\tabcolsep}{8pt}
\begin{tabular}{l|llll}
Case         & Fluxes & $T_i/T_0$ & $p_i / n_0 T_0$ & $c_l$ \\ \hline \\[-10pt]

$\nabla T_i$ & \begin{tabular}[c]{@{}l@{}}Thermo-diffusion,\\ Heat conduction\end{tabular} & $Q^{\text{erf}}(z\vert  1, 0.1 , L)$ & 1 (uniform) & \begin{tabular}[c]{@{}l@{}}Uniform such that \\  DT: $n_D/n_i = n_T/n_i=50\%$,\\ DC: $n_D/n_i=0.975$, $n_C/n_i = 0.025$\end{tabular} \\[10pt]
$\nabla p_i$ & \begin{tabular}[c]{@{}l@{}}Baro-diffusion,\\ Baro-heat flow\end{tabular} & 1 (uniform) & $Q^{\text{erf}}(z\vert  1, 0.5 , L)$ & \begin{tabular}[c]{@{}l@{}} Uniform such that \\ DT: $n_D/n_i = n_T/n_i=50\%$,\\ DC: $n_D/n_i=0.975$, $n_C/n_i = 0.025$\end{tabular} \\[10pt]
$\nabla c$   & \begin{tabular}[c]{@{}l@{}}Classical diffusion,\\ Dufour heat flow\end{tabular} & 1 (uniform) & 1 (uniform) & $Q^{\text{erf}}(z\vert  0.99,0.01,L)$
\end{tabular}}
\caption{Summary of profiles of multi-species cases.}
\label{table:profiles}
\end{table*}
\end{widetext}

\newpage
\bibliography{bib}% Produces the bibliography via BibTeX.

\end{document}